\documentclass[a4paper,11pt]{article}
\usepackage{pos}
\usepackage{setspace}

\renewenvironment{thebibliography}[1]{
\begin{oldthebibliography}{#1}
  \setlength{\itemsep}{0em}
  \setlength{\parskip}{0em}
}
{
  \end{oldthebibliography}
}

\title{On Numerical Simulations of Intergalactic Electromagnetic Cascades with Lorentz Invariance Violation}
 \ShortTitle{Simulations of Intergalactic Electromagnetic Cascades with LIV}

\author*[a,b]{Andrey Saveliev}
\author[c,d]{Rafael {Alves Batista}}

\affiliation[a]{Immanuel Kant Baltic Federal University, \\
  Ul.~A.~Nevskogo 14, Kaliningrad, Russia}

\affiliation[b]{Lomonosov Moscow State University,\\
GSP-1, Leninskiye Gory 1-52, Moscow, Russia}

\affiliation[c]{Instituto de F{\'i}sica Te{\'o}rica UAM-CSIC,\\
C/ Nicol{\'a}s Cabrera 13-15, 28049 Madrid, Spain}

\affiliation[d]{Departamento de F{\'i}sica Te{\'o}rica, Universidad Autónoma de Madrid\\
M-15, 28049, Madrid, Spain}

\emailAdd{andrey.saveliev@desy.de}
\emailAdd{rafael.alvesbatista@uam.es}

\abstract{Lorentz invariance violation (LIV) is a proposed phenomenon where Lorentz symmetry is violated at high energies, potentially affecting particle dynamics and interactions. We use numerical simulations with the CRPropa framework to investigate LIV in gamma-ray-induced electromagnetic cascades, specifically studying how it impacts cascading electrons and photons undergoing pair production and inverse Compton scattering. Our detailed analysis of the simulation results, compared with existing theoretical models, reveals that LIV can significantly alter the behavior of both components of the cascade, photons and electrons, resulting in specific signatures in measured fluxes that could be observed in high-energy gamma-ray observations. These insights are crucial for ongoing searches for LIV and for the development of theoretical models incorporating LIV effects.}

\ConferenceLogo{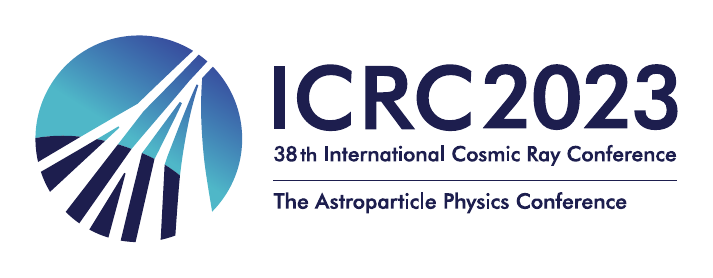}

\FullConference{%
38th International Cosmic Ray Conference (ICRC2023)\\
  26 July - 3 August, 2023\\
  Nagoya, Japan}

\begin{document}
\maketitle

\section{Introduction}

The Standard Model of particle physics (SM) is a highly successful theory that elegantly describes the electromagnetic, weak, and strong interactions \cite{Gaillard:1998ui,CottinghamSM}. It has been incredibly effective in explaining a wide range of physical phenomena. However, there remain several unresolved issues, both from experimental observations and theoretical considerations, that the SM fails to account for. These include such phenomena as Dark Matter, Dark Energy, neutrino masses, and the hierarchy problem \cite{Golling:2016gvc}. 

One of the most fundamental challenges in physics is the unification of the SM and gravity, which is one of the remaining fundamental interactions in nature. Numerous theories have been proposed to address this issue, including Quantum Gravity (QG) \cite{Addazi:2021xuf}, which aims to bridge the gap between the quantum field theory approach of the SM and the differential geometry framework used in General Relativity (GR). GR itself faces several issues, such as singular solutions and the Black Hole information loss problem. Therefore, it is crucial to establish a connection between these theories in order to potentially resolve these problems and gain a deeper understanding of the fundamental forces in the universe.

However, QG is expected to be measurable directly only at the Planck Scale, i.e.~either at distances of the order of the Planck Length, given by $\lambda_{\rm Pl} \simeq 1.62 \times 10^{-35}$ m, or at energies of the order of the Planck Mass, $M_{\rm Pl} \simeq 1.22 \times 10^{28}$ eV. This is 15 orders of magnitude larger than what the Large Hadron Collider can probe with its design center of mass energy of 14 TeV. But if QG is tested specifically, then also other quantities can become important, such as the energy of the particle in some frame or a cosmological propagation distance. These quantities can be large enough to effectively offset the Planck suppression to a physical observable, such that very small corrections are magnified. The most promising candidates for tests of QG \cite{Addazi:2021xuf} are Ultra High Energy Cosmic Rays (UHECR) \cite{Mattingly:2005re,Bietenholz:2008ni,Saveliev:2011vw} and very high energy (VHE) gamma rays \cite{Mattingly:2005re,Galaverni:2008yj,Martinez-Huerta:2020cut}, since they represent the most energetic particles which have been ever observed, their energies reaching values above $10^{20}$ eV.

Specifically, there has been a proposal suggesting that QG could potentially lead to a phenomenon called Lorentz invariance violation (LIV) \cite{Sotiriou:2009bx}. This idea suggests that LIV may arise from the quantum fluctuations of spacetime itself. However, the existence or absence of LIV cannot be definitively proven at this point, as there is no compelling evidence supporting either case. Additionally, it is not possible to completely exclude the possibility of LIV because the Lorentz group, which describes the symmetries of spacetime, is non-compact. Therefore, it is conceivable that LIV could manifest at energies beyond the current limits of observation. In such scenarios, various theoretical approaches, such as effective field theories and SM extensions \cite{Colladay:1998fq}, propose that fundamental aspects of physics, such as dispersion relations and hence threshold energies for different reactions, could undergo modifications. By carefully interpreting relevant observations, these changes may be detected and studied.

In this work we are investigating the consequences of introducing a modified dispersion relation. In particular, we consider so-called electromagnetic cascades which propagate in intergalactic space \cite{Batista:2021rgm}. This describes the process when a VHE gamma-ray photon is emitted by an energetic astrophysical object (e.g.~a blazar or a gamma-ray burst) and then interacts with a low-energy photon of a background photon field, producing an electron/positron pair. Such a charged lepton then upscatters a low-energy photon via Inverse Compton (IC) scattering, resulting in a high energy photon which then can restart the whole process which continues until the particle energy drops below the threshold of the corresponding reaction.

LIV may result in a significant modification of these processes and hence impact the corresponding observations. In particular, \cite{Vankov:2002gt} showed that LIV might dramatically increase the interaction length of pair production for energies above $100$~TeV and therefore suppress the cascade development. On the other hand, LIV might also imply that the photons' speed is energy-dependent, thus resulting in energy-dependent time delays \cite{Martinez-Huerta:2020cut}.

In this work we carry out simulations using the CRPropa code of the propagation of electromagnetic cascades including the modified dispersion relations for VHE photons and electrons for their propagation. We then analyze the influence of the corresponding LIV parameters on the possible observables, such as the measured gamma-ray spectra. To do so, we first describe the formalism of the propagation of photons and electrons/positrons and its modifications due to the modified dispersion relations in Sec.~\ref{sec:ECLIV}. Then, in Sec.~\ref{sec:Simulations}, we describe the setup for the simulations of electromagnetic cascades with LIV and present the corresponding results. Finally, in Sec.~\ref{sec:Discussion}, we discuss our results, while drawing the conclusions from these results and giving an outlook for future investigations we plan to carry out for this topic in Sec.~\ref{sec:Conclusions}.

\section{Particle Propagation with Modified Dispersion Relations} \label{sec:ECLIV}

The aforementioned dispersion relations relevant for electromagnetic cascades are in general given by \cite{Terzic:2021rlx}
\begin{eqnarray}
&E_{e}^{2} = p_{e}^{2} \left[ 1 + \frac{m_{e}^{2}}{p_{e}^{2}} + \sum_{j=0}^{N}\chi_{j}^{e} \left(\frac{p_{e}}{M_{\rm Pl}}\right)^{j} \right] \,, \label{EeDispFull} \\
&E_{\gamma}^{2} = k_{\gamma}^{2}\left[ 1 + \sum_{j=0}^{N}\chi_{j}^{\gamma} \left(\frac{k_{\gamma}}{M_{\rm Pl}}\right)^{j} \right] \label{EgammaDispFull}
\end{eqnarray}
for electrons and photons, respectively. Here, $m_{e}$ is the electron mass, while $p_{e}$ and $k_{\gamma}$ is the electron and photon momentum, respectively. On the other hand, $\chi_{j}^{\gamma}$ and $\chi_{j}^{e}$ are the corresponding LIV parameters of the $j$th order. 

Usually one assumes that one of the terms of the sums (of order $n = 0, 1, 2, ... $) present in Eqs.~(\ref{EeDispFull}) and (\ref{EgammaDispFull}) is dominant, such that these equations %and (\ref{epsilonDispFull}) 
may be reduced to
\begin{eqnarray}
&E_{e}^{2} = p_{e}^{2} \left[ 1 + \frac{m_{e}^{2}}{p_{e}^{2}} + \chi_{n}^{e} \left(\frac{p_{e}}{M_{\rm Pl}}\right)^{n} \right] \,, \label{EeDisp} \\
&E_{\gamma}^{2} = k_{\gamma}^{2}\left[ 1 + \chi_{n}^{\gamma}\left(\frac{k_{\gamma}}{M_{\rm Pl}}\right)^{n} \right]\,, \label{EgammaDisp} 
\end{eqnarray}
respectively. In case of $\xi_{n}^{i} < 0$ the corresponding LIV scenario is also called subluminal, while for $\xi_{n}^{i} > 0$  it is denoted as superluminal.

The consequences of these modifications are twofold \cite{Jacobson:2002hd}: Firstly, the threshold of a given reaction is altered, leading to a modification in the associated propagation length. Secondly, the presence of LIV can enable new reactions that would not be possible within the confines of Lorentz invariance. Examples of such reactions include the spontaneous decay of photons into pairs or photons, as well as the occurrence of the vacuum Cherenkov effect for electrons.

Here, we primarily consider the threshold modification. In order to calculate the threshold of a given reaction, one has to solve the equation 
\begin{equation} \label{ThrCond}
\max_{0 \le \theta \le \pi} s^{\rm in} = \min_{0 \le y \le 1} s^{\rm out} \,,
\end{equation}
where $s^{\rm in}$ and $s^{\rm out}$ is the invariant mass of the incoming and outgoing particles, respectively, given by the symbolic reaction equations
\begin{equation}
\gamma_{\rm VHE} + \gamma_{\rm BP} \rightarrow e^{+}_{\rm VHE} + e^{-}_{\rm VHE}
\end{equation}
for pair production, and
\begin{equation}
e^{\pm}_{\rm VHE} + \gamma_{\rm BP} \rightarrow e^{\pm} + \gamma_{\rm VHE} 
\end{equation}
for Inverse Compton scattering, where in both cases $\gamma_{\rm BP}$ denotes a background photon. The maximum on the left side of the equation should be searched regarding the angle $\theta$ between the propagation directions of the two incoming particles. On the other hand, since the directions outgoing particles for very high energies may be regarded as nearly parallel, the minimum on the right hand side should be searched regarding the fraction $y$ of the total energy carried by one of the outgoing particles.

This results in the background photon threshold energy, $\epsilon_{\rm thr}$, given by
\begin{equation}
\epsilon_{\rm thr} =
\begin{cases}
k_{\gamma} \left[\left( \frac{m_{e}}{k_{\gamma}} \right)^{2} + \frac{1}{4}\left( \frac{\chi_{n}^{e}}{2^{n}} - \chi_{n}^{\gamma} \right) \left(\frac{k_{\gamma}}{M_{\rm Pl}}\right)^{n} \right] & \text{for pair production,}
\\
0 & \text{for inverse Compton scattering,}
\end{cases}
\end{equation}
or, equivalently, in the threshold invariant mass, $s_{\rm thr}$, which may be written as
\begin{equation} \label{sthr}
s_{\rm thr} =
\begin{cases}
k_{\gamma}^{2} \left[ 4 \left(\frac{m_{e}}{k_{\gamma}}\right)^{2} + \frac{\chi_{n}^{e}}{2^{n}} \left(\frac{k_{\gamma}}{M_{\rm Pl}}\right)^{n} \right] & \text{for pair production,}
\\
p_{e}^{2} \left[ \left(\frac{m_{e}}{p_{e}}\right)^{2} + \chi_{n}^{e} \left(\frac{p_{e}}{M_{\rm Pl}}\right)^{n} \right] & \text{for inverse Compton scattering.}
\end{cases}
\end{equation}

Using these results, one then can calculate the mean free path $\lambda_{\rm MFP}$ by \cite{Addazi:2021xuf}
\begin{equation} \label{LIVMFP}
\begin{split}
\lambda_{\rm MFP} = &\frac{1}{2 p_{\rm in}} \int_{s_{\rm thr}}^{\infty} \frac{n_{\rm BP}\left(\frac{s^{*} - p_{\rm in}^{2} \left[ \left(\frac{m_{\rm in}}{p_{\rm in}}\right)^{2} + \chi_{n}^{\rm in} \left( \frac{p_{\rm in}}{M_{\rm Pl}}\right)^{n} \right] }{4 p_{\rm in}}\right)}{\left\{ s^{*} - p_{\rm in}^{2} \left[ \left(\frac{m_{\rm in}}{p_{\rm in}}\right)^{2} + \chi_{n}^{\rm in} \left( \frac{p_{\rm in}}{M_{\rm Pl}}\right)^{n} \right] \right\}^{2}} \\
&\times \int_{p_{\rm in}^{2} \left[ \left(\frac{m_{\rm in}}{p_{\rm in}}\right)^{2} + \chi_{n}^{\rm in} \left( \frac{p_{\rm in}}{M_{\rm Pl}}\right)^{n} \right]}^{s^{*}} \sigma(s) \left\{ s - p_{\rm in}^{2} \left[ \left(\frac{m_{\rm in}}{p_{\rm in}}\right)^{2} + \chi_{n}^{\rm in} \left( \frac{p_{\rm in}}{M_{\rm Pl}}\right)^{n} \right] \right\} \, {\rm d}s \, {\rm d}s^{*}\,,
\end{split}
\end{equation}
where $p_{\rm in}$, $m_{\rm in}$ and $\chi_{n}^{\rm in}$ is the momentum, mass and LIV parameter of the incoming VHE particle, respectively, $n_{\rm BP}(\epsilon)$ is the ambient photon density of the background photon filed, and $\sigma(s)$ is the cross-section of the considered reaction, while $s_{\rm thr}$ is given by Eq.~(\ref{sthr}). 

\begin{figure}[htp] 
  \centering
  \includegraphics[width=0.495\textwidth]{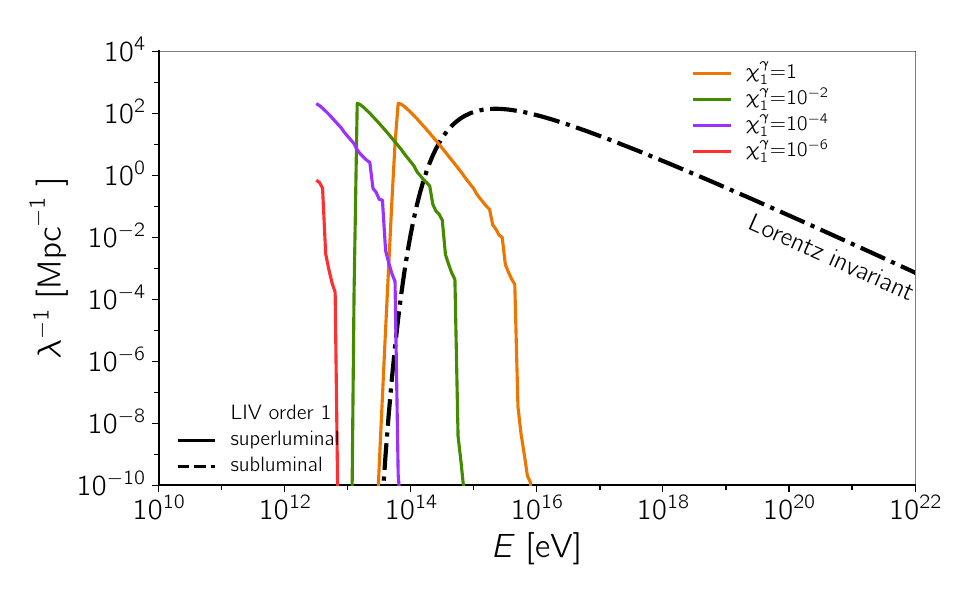}
  \includegraphics[width=0.495\textwidth]{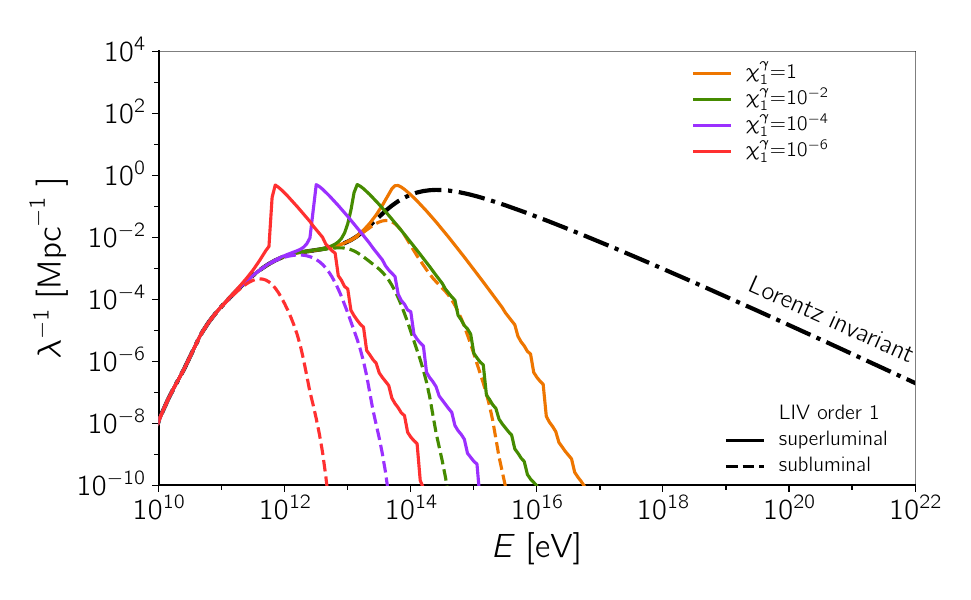}
  \caption{Examples of mean free paths for pair production including LIV calculated according to Eq.~(\ref{LIVMFP}). The left panel corresponds to the CMB, whereas the right one corresponds to the EBL model from \cite{gilmore2012a}. LIV parameters are indicated in the legend.}
  \label{fig:LIVMFP}
\end{figure}

In Fig.~\ref{fig:LIVMFP} we show some examples of mean free paths for pair production by gamma rays according to Eq.~(\ref{LIVMFP}).

\section{Simulations of Electromagnetic Cascades with Lorentz Invariance Violation} \label{sec:Simulations}

We have created an advanced simulation tool called \texttt{LIVPropa} that seamlessly integrates with the CRPropa code~\cite{alvesbatista2016a, AlvesBatista:2022vem}. This plugin enables detailed simulations of gamma-ray propagation while considering LIV. Developed using C++, \texttt{LIVPropa} offers a range of essential features.

One of the key functionalities is the provision of tabulated interaction rate tables tailored for different LIV parameters. To ensure versatility and adaptability, we have also included Python scripts as part of \texttt{LIVPropa}. These scripts facilitate the generation of data for scenarios that may not be readily available within the existing library. This flexibility allows studies of a wide range of LIV scenarios and customization of the simulation parameters to suit the user's specific needs.

\begin{figure}[htb]
    \centering
    \includegraphics[width=0.495\textwidth]{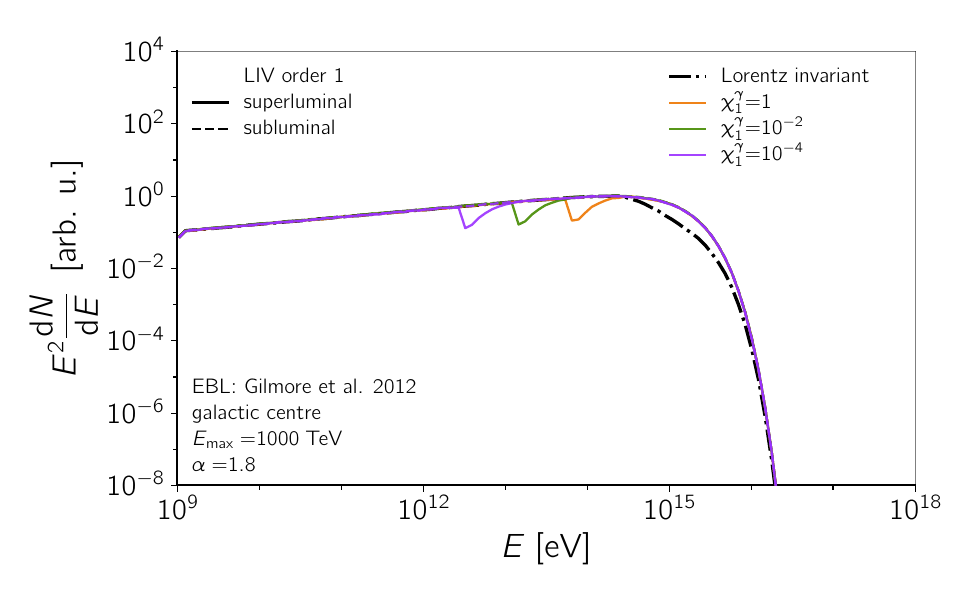}
    \includegraphics[width=0.495\textwidth]{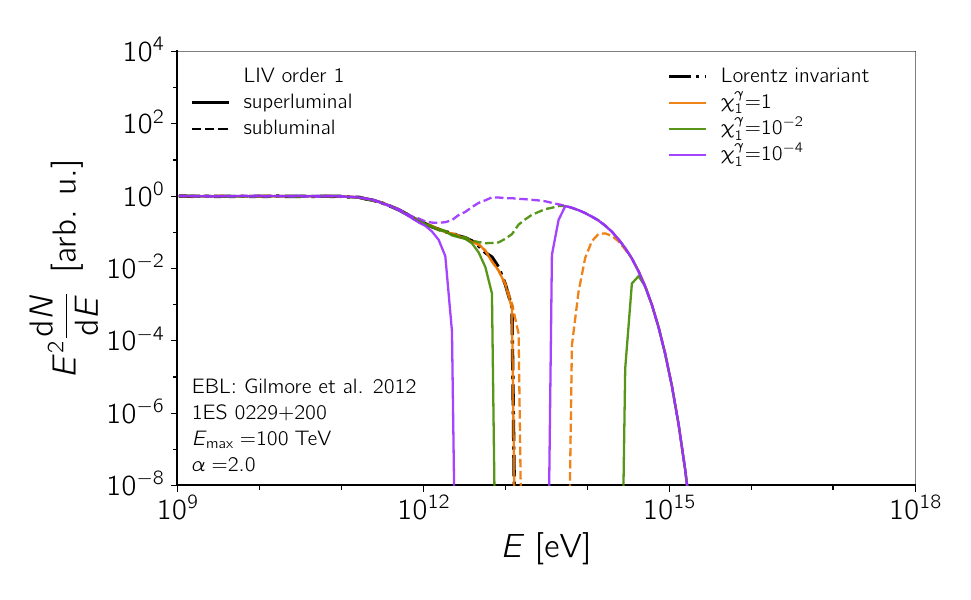}
    \caption{Simulations of gamma-ray propagation consider pair production, including LIV effects (see Eq.~(\ref{LIVMFP})). The left panel corresponds to a cosmologically-distant blazar, whereas the right one is for a gamma-ray source at the Galactic center. The injected spectral parameters assumed are indicated in the panels. The black dot-dashed lines represent the Lorentz invariant scenario.}
    \label{fig:simulationsPP}
\end{figure}

Fig.~\ref{fig:simulationsPP} illustrates the application of the code, considering pair production interactions with the CMB and the EBL. We considered two sources: the extreme blazar 1ES~0229+200 (right panel) and the Galactic center (left panel). Their injection spectrum of both sources follows a power law with index $\alpha$ and an exponential cut-off at an energy $E_\text{max}$.

In the left panel of Fig.~\ref{fig:simulationsPP}, if the Galactic center is a PeVatron, emitting gamma rays with energies up to $\sim 1 \text{PeV}$, characteristic breaks in the spectrum would be visible. These could be observable with with experiments such the Large High Altitude Air Shower Observatory (LHAASO)~\cite{lhasso2019a} or the upcoming Southern Wide-field Gamma-ray Observatory (SWGO)~\cite{swgo2019a}. Note, however, that photons from the Galactic interstellar radiation field also act as targets, potentially modifying the observed fluxes. Nevertheless, the qualitative arguments presented above would still hold.

The right panel of Fig.~\ref{fig:simulationsPP} is very instructive to understand the effect of LIV on the spectra of gamma-ray sources. A suppression due to the EBL is visible at energies around $\sim 1 \; \text{TeV}$. While the exact threshold for this interaction to occur would imply an earlier onset of suppression in the superluminal case, the flux would quickly recover at higher energies, following the behavior of the mean free paths shown in Fig.~\ref{fig:LIVMFP}. In the subluminal case, on the other hand, the behavior closely resembles that of the Lorentz invariant scenario up to an energy related to the energy scale of LIV. Beyond this energy, the flux increases again as a consequence of the large mean free path.

\section{Discussion} \label{sec:Discussion}

For now, \texttt{LIVPropa} takes into account effects related to the modified photon dispersion relation of pair production. However, we are actively working on implementing LIV for inverse Compton scattering as well. This enhancement will enable us to conduct detailed three-dimensional studies of electromagnetic cascades \cite{Batista:2021rgm} that incorporate intergalactic magnetic fields and expand the scope of our simulations.

It is important to note that the phenomenology of LIV encompasses more than just simple modifications of interaction thresholds, which have been the primary focus thus far. Within the framework of a single effective field theory, new processes that are typically forbidden in the Lorentz invariant case begin to emerge. Notable examples include photon decay ($\gamma \rightarrow e^{+} + e^{-}$), photon splitting ($\gamma \rightarrow \gamma + \gamma + \gamma$) and vacuum Cherenkov radiation ($e^{\pm} \rightarrow e^{\pm} + \gamma$) \cite{Jacobson:2002hd, gelmini2005a}. Previous investigations utilizing observations of objects like the Crab nebula at TeV energies have placed constraints on LIV in the QED sector through searches for these effects \cite{astapov2019a, satunin2019a}.

In its current implementation, \texttt{LIVPropa} does not account for time delays resulting from LIV. However, this consideration is mandatory when constraining LIV through time-of-flight measurements~\cite{jacob2008a}. Such strategies have been commonly employed in the search for LIV using gamma-ray observations of flaring blazars with imaging air Cherenkov telescopes (IACTs)~\cite{hess2011a, hess2019a, magic2020c}. Future facilities like the Cherenkov Telescope Array (CTA)~\cite{cta2019a, cta2021b} and the ASTRI Mini-Array~\cite{vercellone2022a} are expected to provide more precise measurements, leading to improved constraints. Consequently, accurate simulation tools like \texttt{LIVPropa} will play a crucial role in predicting the signals expected by these advanced instruments.

\section{Conclusions and Outlook} \label{sec:Conclusions}

In this work we presented the results of Monte Carlo simulations of gamma-ray propagation considering Lorentz invariance violation. To this end, we developed a new software, \texttt{LIVPropa}, which interfaces with the CRPropa code. For now, we implemented LIV for pair production interactions, but this will soon be extended to include inverse Compton scattering.

Ultimately, our goal is to go beyond the usual approach of computing LIV effects by simply folding in the optical depth of the photon field. Instead, we aim to offer a versatile and robust tool for performing Monte Carlo simulations that enable the inclusion of various other relevant factors. These include, for instance, the contributions from Compton-produced photons, the influence of magnetic fields, among others.

\section*{Acknowledgements}

The work of AS is supported by the Russian Science Foundation under grant no.~22-11-00063. RAB is funded by the ``la Caixa'' Foundation (ID 100010434) and the European Union's Horizon~2020 research and innovation program under the Marie Skłodowska-Curie grant agreement No~847648, fellowship code LCF/BQ/PI21/11830030.

% \footnotesize


\begin{thebibliography}{10}
% \setstretch{0.9}

\bibitem{Gaillard:1998ui}
M.~K. Gaillard, P.~D. Grannis, and F.~J. Sciulli.
\newblock {The Standard Model of Particle Physics}.
\newblock {\em Rev. Mod. Phys.}, 71:S96--S111, 1999.

\bibitem{CottinghamSM}
W.~T. Cottingham and D.~A. Greenwood.
\newblock {\em {An introduction to the Standard Model of Particle Physics}}.
\newblock Cambridge University Press, 2023.

\bibitem{Golling:2016gvc}
T.~Golling et~al.
\newblock {Physics at a 100 TeV pp Collider: Beyond the Standard Model
  phenomena}.
\newblock {\em CERN Yellow Rep.}, 3:441--634, 2017.

\bibitem{Addazi:2021xuf}
A.~Addazi et~al.
\newblock {Quantum Gravity phenomenology at the dawn of the multi-messenger era -- a review}.
\newblock {\em Prog. Part. Nucl. Phys.}, 125:103948, 2022.

\bibitem{Mattingly:2005re}
D.~Mattingly.
\newblock Modern tests of Lorentz invariance.
\newblock {\em Living Rev. Rel.}, 8:5, 2005.

\bibitem{Bietenholz:2008ni}
W.~Bietenholz.
\newblock Cosmic rays and the search for a Lorentz invariance violation.
\newblock {\em Phys. Rep.}, 505:145--185, 2011.

\bibitem{Saveliev:2011vw}
A.~Saveliev, L.~Maccione, and G.~Sigl.
\newblock Lorentz invariance violation and chemical composition of ultra high energy cosmic rays.
\newblock {\em JCAP}, 03:046, 2011.

\bibitem{Galaverni:2008yj}
M.~Galaverni and G.~Sigl.
\newblock Lorentz violation and ultrahigh-energy photons.
\newblock {\em Phys. Rev. D}, 78:063003, 2008.

\bibitem{Martinez-Huerta:2020cut}
H.~Mart\'\i{}nez-Huerta, R.~G. Lang, and V.~de~Souza.
\newblock Lorentz invariance violation tests in astroparticle physics.
\newblock {\em Symmetry}, 12(8):1232, 2020.

\bibitem{Sotiriou:2009bx}
T.~P. Sotiriou, M.~Visser, and S.~Weinfurtner.
\newblock {Quantum Gravity without Lorentz invariance}.
\newblock {\em J. High Energy Phys.}, 10:033, 2009.

\bibitem{Colladay:1998fq}
D.~Colladay and V.~A. Kostelecky.
\newblock Lorentz violating extension of the Standard Model.
\newblock {\em Phys. Rev. D}, 58:116002, 1998.

\bibitem{Batista:2021rgm}
R.~{Alves Batista} and A.~Saveliev.
\newblock {The gamma-ray window to intergalactic magnetism}.
\newblock {\em Universe}, 7(7):223, 5 2021.

\bibitem{Vankov:2002gt}
H.~Vankov and T.~Stanev.
\newblock Lorentz invariance violation and the {QED} formation length.
\newblock {\em Phys. Lett. B}, 538:251--256, 2002.

\bibitem{Terzic:2021rlx}
T.~Terzi\'c, D.~Kerszberg, and J.~Stri\v{s}kovi\'c.
\newblock {Probing Quantum Gravity with Imaging Atmospheric Cherenkov
  Telescopes}.
\newblock {\em Universe}, 7(9):345, 2021.

\bibitem{Jacobson:2002hd}
T.~Jacobson, S.~Liberati, and D.~Mattingly.
\newblock {Threshold effects and Planck scale Lorentz violation: Combined constraints from high-energy astrophysics}.
\newblock {\em Phys. Rev. D}, 67:124011, 2003.

\bibitem{gilmore2012a}
R.~C. {Gilmore}, R.~S. {Somerville}, J.~R. {Primack}, and
  A.~{Dom{\'{\i}}nguez}.
\newblock {Semi-analytic modelling of the extragalactic background light and
  consequences for extragalactic gamma-ray spectra}.
\newblock {\em Mon. Not. R. Astron. Soc.}, 422:3189--3207, June 2012.

\bibitem{alvesbatista2016a}
R.~{Alves Batista} et~al.
\newblock {CRPropa 3 -- a public astrophysical simulation framework for
  propagating extraterrestrial ultra-high energy particles}.
\newblock {\em JCAP}, 05:038, 2016.

\bibitem{AlvesBatista:2022vem}
R.~Alves~Batista et~al.
\newblock {CRPropa} 3.2 -- an advanced framework for high-energy particle
  propagation in extragalactic and galactic spaces.
\newblock {\em JCAP}, 09:035, 2022.

\bibitem{lhasso2019a}
{LHAASO Collaboration}.
\newblock The Large High Altitude Air Shower Observatory {(LHAASO)} science
  white paper.
\newblock 2019.

\bibitem{swgo2019a}
{SWGO Collaboration}.
\newblock Science case for a wide field-of-view very-high-energy gamma-ray
  observatory in the southern hemisphere.
\newblock page arXiv:1902.08429, 2019.

\bibitem{gelmini2005a}
G.~{Gelmini}, S.~{Nussinov}, and C.~E. {Yaguna}.
\newblock {On photon splitting in theories with Lorentz invariance violation}.
\newblock {\em JCAP}, 2005(6):012, June 2005.

\bibitem{astapov2019a}
K.~{Astapov}, D.~{Kirpichnikov}, and P.~{Satunin}.
\newblock {Photon splitting constraint on Lorentz invariance violation from
  Crab Nebula spectrum}.
\newblock {\em JCAP}, 2019(4):054, 2019.

\bibitem{satunin2019a}
P.~{Satunin}.
\newblock {New constraints on Lorentz Invariance violation from Crab Nebula
  spectrum beyond 100 TeV}.
\newblock {\em Eur. Phys. J. C}, 79(12):1011, December 2019.

\bibitem{jacob2008a}
U.~{Jacob} and T.~{Piran}.
\newblock Lorentz-violation-induced arrival delays of cosmological particles.
\newblock {\em JCAP}, 2008(1):031, 2008.

\bibitem{hess2011a}
{H.E.S.S. Collaboration}.
\newblock {Search for Lorentz Invariance breaking with a likelihood fit of the
  PKS 2155-304 flare data taken on MJD 53944}.
\newblock {\em Astropart. Phys.}, 34(9):738--747, April 2011.

\bibitem{hess2019a}
{H.E.S.S. Collaboration}.
\newblock {The 2014 TeV $\gamma$-ray flare of Mrk 501 seen with H.E.S.S.:
  temporal and spectral constraints on Lorentz invariance violation}.
\newblock {\em Astrophys. J.}, 870(2):93, January 2019.

\bibitem{magic2020c}
{MAGIC Collaboration}.
\newblock Bounds on Lorentz invariance violation from magic observation of
  {GRB} 190114c.
\newblock {\em Phys. Rev. Lett.}, 125(2):021301, 2020.

\bibitem{cta2019a}
{CTA Consortium}.
\newblock {\em Science with the Cherenkov Telescope Array}.
\newblock World Scientific, 2019.

\bibitem{cta2021b}
{CTA Consortium}.
\newblock Sensitivity of the Cherenkov Telescope Array for probing cosmology
  and fundamental physics with gamma-ray propagation.
\newblock {\em JCAP}, 02:048, 2021.

\bibitem{vercellone2022a}
S.~{Vercellone} et~al.
\newblock {ASTRI Mini-Array core science at the Observatorio del Teide}.
\newblock {\em J. High Energy Phys.}, 35:1--42, 2022.

\end{thebibliography}
\end{document}